%% file: CAMERA_READY_NetSciCom2017.tex
\documentclass[10pt, conference, letterpaper]{IEEEtran}
\usepackage{array}
\usepackage{mdwmath}
\usepackage{mdwtab}
\usepackage{graphicx}
\usepackage{float}
\usepackage{stfloats}
\usepackage{lipsum}
\usepackage{tabularx}
\usepackage[tight,footnotesize]{subfigure}
\usepackage{caption}
\usepackage{epstopdf}
\usepackage{amsmath}
\usepackage{amssymb}
\usepackage{paralist}
\usepackage{mathrsfs}
\usepackage{url}
\usepackage{mathbbol}
\usepackage{times}

\usepackage{booktabs}
\usepackage{verbatim}

\usepackage{bm}			% bold math even for greek letters

\usepackage{amsthm}

\newtheorem{definition}{Definition}

\usepackage[table]{xcolor}
\usepackage{color,colortbl}

\definecolor{Gray}{gray}{0.9}

 \usepackage[normalem]{ulem}
 
\begin{document}

\title{Re-mapping the Internet: Bring the IXPs into Play \\ {\Large \url{www.inspire.edu.gr/ixp-map} }  }
\author{Pavlos Sermpezis\\FORTH, Greece\\sermpezis@ics.forth.gr \and
		George Nomikos\\FORTH, Greece\\gnomikos@ics.forth.gr \and
		Xenofontas Dimitropoulos\\FORTH / University of Crete, Greece\\fontas@ics.forth.gr }

\maketitle

\begin{abstract}
\input{Abstract}
%\blfootnote{This work was funded by the European Research Council via Grant Agreement no. 338402, project ``NetVolution: Evolving Internet Routing''.}
\end{abstract}

\section{Introduction}\label{sec:intro}
\input{Introduction}

\section{The IXP-based Model}\label{sec:model}
\input{Model}

\section{Dataset Analysis}\label{sec:dataset-analysis}
\input{Measurement}

\section{Related Work}\label{sec:related}
\input{Preliminaries_pavlos}	

\section{Conclusion}\label{sec:conclusion}
\input{Conclusion}
\vspace{\baselineskip}
\noindent\textbf{Acknowledgements.} This work has been funded by the European Research Council grant agreement no. 338402.
%\vspace{-0.5\baselineskip}

\bibliographystyle{IEEEtran}	
%\small

\end{document}

%% file: Abstract.tex
%The knowledge of the Internet topology is of high importance in designing networks and architectures, optimizing services, evaluating performance, or network economics. Interconnections between Autonomous Systems (ASes), routers, and their points of presence (PoPs), have been measured, and their properties, like connectivity or link capacity, have been analyzed. However, existing modeling approaches have some serious shortcomings, related to ease, accuracy and completeness of measurements, and limited applicability to emerging research areas. To this end, in this paper, we propose a novel approach towards modeling the \textit{inter-domain} Internet topology. Motivated by the recent interest in the Internet eXchange Points (IXPs) and the increasing importance of their role in the Internet ecosystem, we introduce a network graph model having as main components the IXPs and their AS memberships. The proposed IXP-based model aims to complement previous modeling efforts, shed light on unexplored characteristics of the Internet topology, and support new research directions. We also collect Internet connectivity data, build an IXP-based graph, analyze main topological properties, and discuss application-related issues.
The Internet topology is of high importance in designing networks and architectures, evaluating performance, and economics. Interconnections between domains (ASes), routers, and points of presence (PoPs), have been measured, analyzed, and modeled. However, existing models have some serious shortcomings, related to ease, accuracy and completeness of measurements, and limited applicability to emerging research areas. To this end, in this paper, we propose a novel approach towards capturing the \textit{inter-domain} Internet topology. Motivated by the recent interest in the Internet eXchange Points (IXPs), we introduce a network graph model based on IXPs and their AS memberships. The proposed model aims to complement previous modeling efforts, shed light on unexplored characteristics of the Internet topology, and support new research directions. We also collect and make available Internet connectivity data, analyze main topological properties, and discuss application-related issues.

%% file: Introduction.tex
%\textbf{What is Internet topology, how is it defined?}
%
The Internet is composed of Autonomous Systems (ASes), like ISP networks, CDNs, university networks, located all over the world. 
%The interconnection of ASes over the Internet gives to their users the ability to access information and services, or exchange traffic. 
To connect with each other, ASes establish physical links and use routers, located in different points-of-presence (PoPs). Connections can be either private (between a pair of ASes) or hosted by an Internet eXchange Point (IXP). %, where an AS can connect to more than one ASes using the same equipment, decreasing thus its costs and increasing its accessibility \cite{augustin2009ixps}. 
This Internet ecosystem (of ASes, routers, links, PoPs, IXPs, etc.) is highly dynamic, and its evolution is driven by both technological and economic factors~\cite{griffin2002stable, pch-economics}. Changes in the network topology are frequent and are done in a distributed way; ASes decide individually if and how to add, remove, or modify (physical or logical) links to other ASes. 

Knowing the current topology of the Internet can help operators to design, configure, or optimize their network% and connectivity infrastructure
~\cite{cisco2005}. However, since the Internet is not centrally controlled, and ASes do not use to publish their private network structure and peering relationships, it becomes difficult (if not impossible) to have a centralized/global view of the Internet topology~\cite{roughan201110}.

%\textbf{Literature on Internet topology measurements and analysis.}
%
Only partial knowledge can be obtained by measuring the Internet from the edge, or some vantage points (using traceroutes, looking glass servers, etc.)~\cite{ager2012anatomy,augustin2009ixps}. The collected data can then be used to build a \textit{network graph}, which is a common modeling approach that facilitates analysis of the network characteristics. Graph nodes/vertices often correspond to ASes, routers, or PoPs (resulting to an \textit{AS-}~\cite{routeviews}, \textit{router-}~\cite{pansiot1998routes}, or \textit{PoP-} \cite{knight2011internet} \textit{level} topology, respectively), while edges denote a physical (or logical) link between nodes.
%Representing the Internet topology with a graph is a common modeling approach that can often facilitate analysis and detection of main structural and functional characteristics of the network. The graph vertices can correspond either to ASes, routers, or PoPs (resulting to an \textit{AS-} \red{[..]}, \textit{routing-} \red{[..]}, or \textit{PoP-} \red{[..]} \textit{level} topology, respectively), while the edges denote that there exist physical connections (e.g., optical fiber links) between the nodes/vertices.

%% REMOVE due to space limitation
%Based on properties observed in these real -but usually inclomplete- graphs, a number of synthetic models have been also proposed \red{[..,..]}. Their goals are to generate synthetic topologies, which can be a very useful tool for the performance evaluation of different protocols/mechanisms under realistic conditions, etc. \red{[..]}, or to infer a more complete Internet graph \red{[..]}.

%\textbf{Shortcomings of previous approaches.}
%
Despite their usefulness (providing an initial view of the network structure, in simulation studies, etc.), existing network models are in some cases insufficient to provide a thorough view of the Internet topology, or lead researchers to correct conclusions and insights~\cite{Willinger-Internet-Topology-ebook-2013}. This is mainly due to measurement limitations, or absence of verification (``ground truth'') for the proposed network models~\cite{roughan201110,pansiot1998routes,zhao2001analysis,shavitt2010structural}.

To (i) \textit{complement} previous models, (ii) overcome some of their shortcomings, and (iii) keep up with the recent Internet research directions, in this paper, we propose a new, fundamentally different approach for mapping the \textit{inter-domain} Internet topology. At the heart of our model are the \textit{Internet eXchange Points (IXPs)}: the facilities where many ASes meet and exchange traffic. Our selection for an IXP-based model is motivated by the fact that IXPs (a) reside in the core of the Internet, and (b) play an increasingly important role in the Internet's connectivity (e.g., peering links~\cite{ager2012anatomy,augustin2009ixps}) and structure (e.g., flattening of the topology~\cite{dhamdhere2010internet,gregori2011impact}).%, which are the two main factors that motivate our selection for an IXP-based model.

Specifically, we define the \textit{IXP bipartite graph}, a graph model that consists of two disjoint sets of vertices, i.e., IXPs and ASes. An edge between an IXP-AS pair is defined when the AS is member of the IXP; edges between AS-AS or IXP-IXP pairs are not allowed. The importance and usefulness of the IXP bipartite graph lies mainly on the following directions:

\begin{itemize}
\item The information needed to build the IXP bipartite graph (i.e., IXP memberships) is published by the IXPs themselves (which have incentives to do so), and is easily accessible~\cite{pdb,pch,euroix}. Thus, the proposed representation of the Internet does not rely on measurements, which are a main cause of inaccuracies in previous models.
%Thus, a number of problems of previous models related to measurement inaccuracies, can be overcome.

%The information needed to build the IXP bipartite graph (i.e., IXP memberships) is publicly available and easily accesible~\cite{}. Moreover, IXPs have incentives to publish this information, in order to increase their popularity, attract clients, etc.~\cite{}. Thus, a number of problems of previous models related to measurement inaccuracies, can be overcome.

%a number of problems of previous models related to measurement inaccuracies, can be overcome.

\item A lot of recent studies focus on IXPs and their role in the Internet~\cite{ager2012anatomy,augustin2009ixps,dhamdhere2010internet,gregori2011impact,CXP,Gupta-SDX-CCR-2014}. The IXP-based model can become a useful tool in IXP-related research. Using the IXP bipartite graph, instead of or in conjunction with the AS/router/PoP-level graphs, can (i) shed light on further characteristics of the Internet topology, and (ii) enable new research on areas like inter-domain routing, or network economics.% (as we show in Section~\ref{sec:dataset-analysis}).

%\item A lot of recent studies focus on IXPs and their role in the Internet~\cite{ager2012anatomy,augustin2009ixps,dhamdhere2010internet,gregori2011impact,CXP}. The IXP-based model can serve as \textit{a common framework}, which is currently missing, and become a useful tool in IXP-related research.
% 
%\item Using the IXP bipartite graph, instead of or in conjunction with the AS/router/PoP-level graphs, can (i) shed light on further characteristics of the Internet topology, and (ii) enable new research on areas like inter-domain routing, or network economics (as we show in Section~\ref{sec:dataset-analysis}).

%@Pavlos: it wasn't clear to me what you mean with "common framework" Maybe explain a bit why a common framework is useful. We dont touch this much later on. Otherwise maybe merge the two items (this and the next one) and remove the "common framework". Fontas

\end{itemize}

The remainder of the paper is structured as follows. We first introduce the IXP-based model, and discuss its main characteristics and applicability to use cases (Section~\ref{sec:model}). Then, we build the IXP bipartite graph from collected data (which we make publicly available~\cite{ixp-bg-dataset-new}), analyze its main characteristics, and provide application-related insights (Section~\ref{sec:dataset-analysis}). Finally, we discuss related work (Section~\ref{sec:related}) and conclude our paper (Section~\ref{sec:conclusion}).
%highlighting important findings and implications, 

%To present the model and the contributions discussed above, we structure the paper as follows. We first provide some basic information about Internet models and the role and function of IXPs (Section~\ref{sec:preliminaries}). Then, we introduce the IXP-based graphs (Section~\ref{sec:model}) and analyze their main characteristics (Section~\ref{sec:dataset-analysis}). Finally, we demonstrate how our framework and findings can be applied to a number of use cases (Section~\ref{sec:applications}) and conclude our paper (Section~\ref{sec:conclusion}).

%% file: Model.tex
In this section, we introduce our Internet topology model. After providing some basic information about IXPs (Section~\ref{sec:preliminaries-IXPs}), we define the model, the \textit{IXP bipartite graph} and its derivatives (Section~\ref{sec:model-definition}). Then, we discuss its main characteristics (Section~\ref{sec:model-characteristics}), as well as extensions of it (Section~\ref{sec:model-extensions}). Finally, we provide some use cases where applying our model could facilitate solutions for practical problems (Section~\ref{sec:model-applications}).

\subsection{Preliminaries: Internet eXchange Points}\label{sec:preliminaries-IXPs}
Connections between ASes can take place over private links (between a pair of ASes) or Internet eXchange Points. The main functionality of an IXP is equivalent to a {layer-2} switch. An AS \textit{member} of the IXP has a router (or a few routers) connected to the IXP infrastructure, and, thus, is able to establish a connection with \textit{any} other IXP member. The main incentive for an AS to become member of an IXP, is that, using the same equipment (e.g., $1$ router, $1$ physical link), it can connect to many ASes, which leads to cost decrease (compared to private links) and increase of its accessibility.

\begin{figure}
\centering
\includegraphics[width=0.75\linewidth]{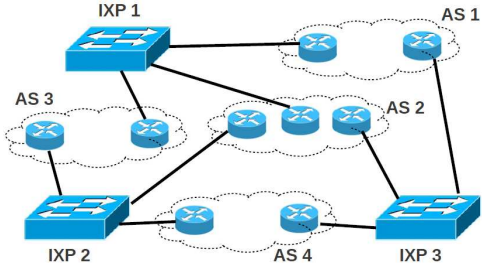}
\caption{Physical topology of a network with $4$~ASes and $3$~IXPs.}
\label{fig:IXP-real-topology}
\end{figure}

Fig.~\ref{fig:IXP-real-topology} shows an example network of $4$ ASes (depicted as clouds) and $3$ IXPs (depicted as switches); this could be a part of the Internet topology. An AS can connect to more than one IXPs, from different locations of its intra-network (e.g., AS-1 connects to IXP-1 and IXP-3 from two different routers).

\subsection{Model Definition}\label{sec:model-definition}

We now define our IXP-based model for the Internet topology, the \textit{IXP bipartite graph} (BG). Let us first denote the set of IXPs and ASes in the Internet as $\mathcal{IXP}$ and $\mathcal{AS}$, respectively. We also denote that an AS $j\in\mathcal{AS}$ is member of an IXP $i\in\mathcal{IXP}$ as $j\rightarrow i$. The IXP BG is defined as follows.

\begin{figure}
\centering
\includegraphics[width=0.75\linewidth]{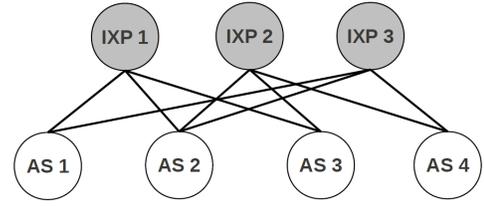}
\caption{IXP bipartite graph (\textit{IXP BG}).}
\label{fig:IXP-bipartite}
\end{figure}

\begin{definition}[IXP bipartite graph]\label{def:bipartite-graph}
The IXP bipartite graph is the undirected graph $\mathcal{BG} = \{\mathcal{V}_{BG},\mathcal{E}_{BG}\}$ whose vertices are the IXPs and ASes of the network:
\[\mathcal{V}_{BG} =\mathcal{IXP} \cup \mathcal{AS}\]
An edge $e_{ij}$ between two vertices $i\in \mathcal{IXP}$ and $j\in\mathcal{AS}$ exists when $j$ is a member of $i$, i.e.,
\[\mathcal{E}_{BG} = \{e_{ij}: i\in\mathcal{IXP} , j\in\mathcal{AS} , j\rightarrow i\}\]
\end{definition}
Under the above definition, edges can exist only between an IXP and an AS; IXP-to-IXP and AS-to-AS edges do not exist in the IXP BG. As a result, the graph of Def.~\ref{def:bipartite-graph} consists of two \textit{disjoint} sets of nodes, the IXPs and the ASes, or, equivalently, it is a \textit{bipartite} graph. In Fig.~\ref{fig:IXP-bipartite} we draw the IXP BG that corresponds to the example network topology of Fig.~\ref{fig:IXP-real-topology}. We denote the nodes with circles (gray and white, for IXPs and ASes, respectively) and the edges with lines.
%\footnote{In practice, there exist private links betweens ASes, and thus AS-to-AS edges exist as well (IXP-to-IXP connections are rare). However, we remind that the focus of our model is different than in previous models (which consider AS-to-AS links), and it is \textit{complementary}, and not a subsitute, to them.}
%\footnote{In practice, there exist private links betweens ASes, and thus AS-to-AS edges exist as well (IXP-to-IXP connections are rare). However, we remind that the goals of our model (see, e.g., analysis of topological characteristics and related applications in Sections~\ref{} and~\ref{}) are different than in previous models (e.g., AS-graph) considering also private AS-to-AS links information; and it is \textit{complementary}, and not a subsitute, to them.}

\underline{Remark:} In practice, AS-to-AS edges exist as well due to private links (IXP-to-IXP connections are rare). However, we remind that the goals of our model (towards the analysis of topology characteristics, or applications; see, e.g., Section~\ref{sec:dataset-analysis}) are different from previous models, like the AS-graphs, which consider also private AS-to-AS links. Moreover, the IXP BG can be used \textit{complementarily} with other models as we discuss later on.

%In cases that a more detailed view of the Internet is needed, the IXP BG can still be used \textit{complementarily} with other models.

%A small graph size is essential for many applications employing online algorithms. When needed, graphs that are derivatives of the IXP BG and have smaller size can be used.
\textbf{Graph Derivatives.} From the basic model of the IXP BG, one can build two derivative graphs. These graphs are simpler, may fit different use cases better, and are of smaller size, a property that is essential for many applications employing online algorithms~\cite{CXP}. Specifically, one can build the \textit{IXP multigraph} (MG). The IXP MG is defined as \textit{the projection of the BG onto the set of the IXP nodes}\footnote{The projection of a bipartite graph onto one set of its nodes $X$, is defined as a graph (with single or multiple edges) whose nodes are in $X$, and two nodes are connected if they have a common neighbor in BG.}. Nodes in IXP MG are the IXPs, i.e., $\mathcal{V}_{MG} = \mathcal{IXP}$, and an edge $e_{ij}$ between two nodes $i,j\in \mathcal{IXP}$ exists when an AS is member of both IXPs $i$ and $j$, i.e.,
\[\mathcal{E}_{MG} = \{e_{ij}: i,j\in\mathcal{IXP} , k\in\mathcal{AS} , k\rightarrow i, k\rightarrow j\}\]
As it can be seen, the IXP MG consists of much less nodes than the BG ($|\mathcal{V}_{MG}| = |\mathcal{IXP}|\ll |\mathcal{V}_{BG}| = |\mathcal{IXP}|+|\mathcal{AS}|$).

\begin{figure}
\centering
\includegraphics[width=0.75\linewidth]{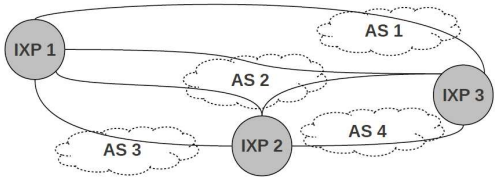}
\caption{IXP multi-graph (\textit{IXP MG}).}
\label{fig:IXP-multi-graph}
\end{figure}

Since more than one ASes can be members at the same two IXPs, the resulting graph is a \textit{multigraph}. In the example of Fig.~\ref{fig:IXP-real-topology}, $AS_{2}$ and $AS_{3}$ are members to both $IXP_{1}$ and $IXP_{2}$. Thus, in Fig.~\ref{fig:IXP-multi-graph}, where we present the corresponding IXP multigraph, the nodes $IXP_{1}$ and $IXP_{2}$ are connected with two edges (through $AS_{2}$ and $AS_{3}$). 

The \textit{AS multigraph} (or, AS MG) can be defined and~represented similarly; we omit the details due to space limitations.

\subsection{Model Characteristics}\label{sec:model-characteristics}
We proceed to discuss issues related to the construction of the graph, how it compares to existing Internet topology models, and
its main characteristics and usefulness.
 
\textbf{Graph Construction.} To build the IXP bipartite graph (BG) we need to know \textit{the set of IXPs and their member ASes}, i.e., the ASes that are physically connected to each IXP. With this information the construction of the IXP BG is straightforward; one needs simply to follow the rules of Def.~\ref{def:bipartite-graph}.%\footnote{Virtual links (e.g., remote AS members) can be considered as well.}

Information about AS memberships is usually \textit{publicly available} at IXPs' websites, or in databases like PeeringDB~\cite{pdb}, PCH~\cite{pch}, or Euro-IX~\cite{euroix}, that bring together aggregate data about IXPs and their members. Moreover, IXPs (as well as  ASes) {\it have incentives to publish this information}, in order to attract peers or clients and increase their popularity~\cite{roughan201110}. This makes the construction of the IXP BG a much simpler task compared to previous modeling approaches, which require complex data collection and processing methods to {\it infer} AS/router/PoP-level links. This information is typically kept private by ISPs, in sharp contrast to our model, for security and competitive strategy reasons.  

Furthermore, measurement methods needed for building AS/router/PoP-level graphs have been extensively reported to suffer from a number of deficiencies (see, e.g., the survey of \cite{Willinger-Internet-Topology-ebook-2013}). Protocols (e.g., BGP messages) or tools (e.g., traceroute) used in these methods, have not been designed for Internet topology measurements. For instance, BGP messages contain only best paths, and, thus, many existing (e.g., backup) links are not included in the AS-graphs~\cite{roughan201110}. In traceroute-based methods, the difficult task of IP alias resolution (i.e., matching IP addresses to routers) can lead to inaccurate router-level graphs~\cite{keys2010internet,pansiot1998routes}. On the contrary, AS membership data are independent of such technological approaches, and avoid problems related to measurements incompleteness/inaccuracy. Indeed, early studies on the correctness of the available IXP membership data are very encouraging~\cite{Kloti-Comparative-IXP-CCR-2016,traixroute}.

%Moreover, the measurement methods used for building AS- / router- / PoP- level graphs has been extensively reported to suffer from a number of deficiencies (\red{[cite at least 3]}, to name a few), resulting to inaccurate and/or incomplete representations of the Internet topology. This is mainly due to the fact that data collection methods are based on specific protocols (e.g., BGP messages) or tools (e.g., traceroute) that have not been designed for Internet topology measurement; thus, there are many limitations when we use them to construct the Internet graph. For instance, BGP messages that are used for the constuction of the AS-graph contain only best paths, and, thus, many existing (e.g., backup) links between ASes are not included~\cite{}. \blue{[at least another example]}. On the contrary, the available AS membership data are independent of such technological approaches and have been shown to have a higher accuracy and completeness (most AS memberships are reported, there exist only a few stale data, etc.) \blue{[is this true? - citation needed]}\footnote{Although some inaccuracy and incompleteness exist also in the AS membership datasets, mainly due to the fact that most data are self-reported, combining information from more than one sources has been shown to improve the quality of the data [cite CCR paper].}.

Summarizing, the IXP BG construction process has the following advantages compared to previous models: (a) ease of collection, 
(b) independence of measurements, and (c) potential for increased accuracy.

\textbf{Role and Use.} The proposed IXP-based model, and its derivatives, map parts of the \textit{inter-domain} Internet topology. Hence, it is conceptually closer to the AS-graph model, rather than the (fine-granular) router/PoP-level models. However, the IXP BG offers a view of the topology from IXPs, whereas the AS-graph is usually built from information collected at the edge~\cite{roughan201110}. Moreover, the edges in the IXP BG are well-defined: they are physical connections between ASes' edge routers and IXPs. In contrast, an AS-graph edge might correspond to more than one physical links located in distant locations, which can create misunderstandings and/or lead to inaccurate conclusions~\cite{roughan201110,Willinger-Internet-Topology-ebook-2013}.

%Since the IXP BG offers a view of the inter-domain topology, as it is seen from the IXPs, it can be useful in research areas, like inter-domain routing~\cite{Peter-One-Tunnel-Sigcomm-2014,Hu-Measurement-ID-Routing-Diversity-2012,CXP}, IXP-related innovation~\cite{Gupta-SDX-CCR-2014,CXP}, or network economics~\cite{Mahajan-Negotiation-based-routing-USENIX-2005}. In Section~\ref{sec:dataset-analysis}, we discuss a number of use cases and demonstrate the applicability of our model. %Furthermore, combining the information provided by the IXP BG and previous models, can support efforts towards a more complete mapping of the AS topology~\cite{dhamdhere2010internet,gregori2011impact}, PoP topology~\cite{giotsasmapping}, etc.

\subsection{Enriching the Graph}\label{sec:model-extensions}
Extra information can be added (e.g., as attributes to nodes or edges) in the model of Section~\ref{sec:model-definition} to extend its applicability. Some examples are presented in the following.

\textbf{Location data.} IXPs are located in one or a few nearby buildings. Therefore, a location attribute (e.g., country, city, or GPS coordinates) can be assigned to an IXP-node~\cite{giotsasmapping}\footnote{ASes might be large networks spanning a large geographical area. Therefore, assigning a location to an AS node is not always possible.}.%, and be used for topology analysis or application-specific purposes., 

\textbf{AS type.} ASes can be classified according to their business type. For example as AS can be an ISP (Internet Service Provider), a CDN (Content Distribution Network) , an education/research organization, an enterprise, etc. Assigning types to AS-nodes, and analyzing the resulting graphs, could help us understand further the Internet structure and the underlying economics, as we show in Section~\ref{sec:dataset-analysis}.

\textbf{IP prefixes.} The IP prefixes (or, \textit{customer cone}) of each AS can be added as an attribute to AS-nodes. This information can then be used in routing problems, but also as an indicator of the AS importance/size~\cite{CXP}, thus allowing a \textit{joint investigation} of the structural and functional Internet characteristics.

\textbf{Peering policies.} Peering agreements (e.g., peer-to-peer, customer-to-provider) between ASes determine if and how they exchange traffic over an IXP fabric. Hence, a ``layer'' containing policy information can be added on top of the IXP BG or the AS MG, (as directed or active/inactive edges, etc.). %This (logical) layer is orthogonal to the structure of the graph (i.e., the physical links) and affects only certain processes taking place over the graph (e.g., a routing algorithm).

\subsection{Applications}\label{sec:model-applications}
Since the IXP BG offers a view of the inter-domain topology, as it is seen from the IXPs, it can be useful in research areas, like inter-domain routing~\cite{Peter-One-Tunnel-Sigcomm-2014,Hu-Measurement-ID-Routing-Diversity-2012,CXP}, IXP-related innovation~\cite{Gupta-SDX-CCR-2014,CXP}, or network economics~\cite{Mahajan-Negotiation-based-routing-USENIX-2005}. 

Solving practical problems relating to selection of IXPs (e.g., based on their topological characteristics), can be facilitated using the IXP BG, by making it easier to apply already used heuristics (based on costs, performance, etc.). Moreover, the IXP BG provides a graph representation of the topology, which makes easier to formalize a number of IXP-related problems as the well-known \textit{set cover} problems~\cite{budget-max-cover,krause2012submodular}. Hence, the large literature on set cover problems can be used towards finding optimal solutions even for complex cases. In the following (as well as in Section~\ref{sec:dataset-analysis}) we discuss some relevant examples\footnote{Expressing the corresponding optimization problems is straightforward, but requires a detailed examination of the problem, to select the optimization variables, constraints, etc., which is beyond the scope of this paper.}.

\textbf{Outsourcing middleboxes} is a novel way to decrease infrastructure, costs, etc., for enterprises~\cite{Sherry-Making-Middleboxes-CCR-2012}. Candidate locations for installing middleboxes are IXPs~\cite{decix-blackholing}, due to their accessibility (for ASes and potential clients), good connectivity (high bandwidth, availability, etc.), and colocation with data centers~\cite{CXP}. Since the traffic of an enterprise needs to be directed to the middlebox and then redirected to its source or destination, IXPs in \textit{central points of the Internet} are expected to be good selections for satisfying QoS constraints (e.g., latency). To this end, a middlebox placement strategy could benefit from the knowledge of the IXP BG. The middlebox/service provider needs to select the set of IXPs, where the equipment will be installed, e.g., having as an objective (a) to cover all its (potential) customers with the minimum deployment cost (e.g., number of sites where middleboxes will be installed, resources per installation site, etc.), or (b) given a cost (CAPEX/OPEX) to cover as many customers as possible. 

\textbf{Remote peering} has been recently appeared as a service in major IXPs~\cite{remote-peering-euro-ix,decix-remote-peering}. In remote peering, a network ``reseller'' (e.g., an ISP) allows ASes to have (virtual) presence to remote IXPs, i.e., to IXPs that are located outside the area that the network of the AS expands. As a result, ASes are not constrained by locality criteria in the selection of a new membership to an IXP. This allows more flexibility in setting up new memberships, and the IXP BG can be used for an optimal selection of the IXP to which an AS will be connected, based on the current availability of options (i.e., resellers for remote peering), connection costs, etc.

\textbf{Deploying a new IXP}, can be considered as an equivalent problem as above, but now from an IXP point of view. A new IXP (or an existing one that expands its infrastructure) combining location information (see, e.g., Section~\ref{sec:model-extensions}) with the IXP-BG, can similarly select the location of a new site deployment based on the ``coverage'' (e.g., how many ASes are located there, and if they are already well connected through other IXPs) in order to maximize its expected revenue.

%% file: Measurement.tex
In this section, we build the IXP BG for an instance of the Internet% (from data collected on Nov. 20th, 2015)
. We first briefly describe the datasets and methodology we use (Section~\ref{sec:datasets}), and then analyze the graph (Section~\ref{sec:analysis}).

%Our findings shed light on the structure and characteristics of the Internet topology. We also highlight important implications, and discuss how the IXP-based model can be a useful tool for applications attracting increasing interest, like, (a)~IXP selection strategies for CDNs and ISPs, (b)~remote peering, (c)~outsourcing middleboxes, (d)~traffic engineering, etc.  

\subsection{Datasets}\label{sec:datasets}
As discussed earlier, a key feature of the proposed model is the \textit{ease} of the graph construction. The only information needed to build the IXP BG is the \textit{IXP memberships}. In our study, we collect these data from the \textsc{PeeringDB} (PDB) \cite{pdb} and \textsc{Packet Clearing House} (PCH)~\cite{pch} websites. %, which provide \textit{publicly available}, \textit{up-to-date} data of ASes connected to IXP fabrics. 
Combining two information sources, gives a more complete graph.%\footnote{Data from the Euro-IX~\cite{TODO} website can increase the completeness of the graph (by a small percentage $\sim15\%$, as shown in~~\cite{TODO,ccr-paper}). However, Euro-IX does not contain information about IXP subnets, complicating thus the necessary merging process. Due to space limitations, we do not consider Euro-IX data in our analysis.}. 

The (simple) methodology we use to merge and pre-process the data is as follows. Since both datasets do not share the same identifiers for the IXPs, we merge the two IXP lists by matching the IXP IP prefixes and names. %After merging, the IXP and AS lists are augmented by \red{X\%} and \red{X\%}, respectively, comparing to the largest individual dataset. 
Then, we filter out the IXPs denoted as \textit{inactive} or \textit{not approved} in PDB and PCH datasets. We also discard AS-IXP edges with inconsistent information (based on IP addresses), and nodes that do not belong to the giant component ($3\%$ and $1\%$ node and edge discards, in total). The final graph is connected, and consists of $504$ IXP and $4,692$ AS nodes, and $14,651$ IXP-AS edges. %\red{What fraction of total ASes and IXPs is included in our dataset?}
Finally, to enrich the basic IXP BG and complement our analysis, we use ASes' prefixes from PDB and PCH, and  ASes' types and peering relationships from CAIDA~\cite{caida}.

To facilitate further analysis of the IXP graphs, we make available the IXP-AS membership datasets and the respective graphs at~\cite{ixp-bg-dataset-new}.

\begin{figure}
\centering
\subfigure[Node degree]{\includegraphics[scale=0.23]{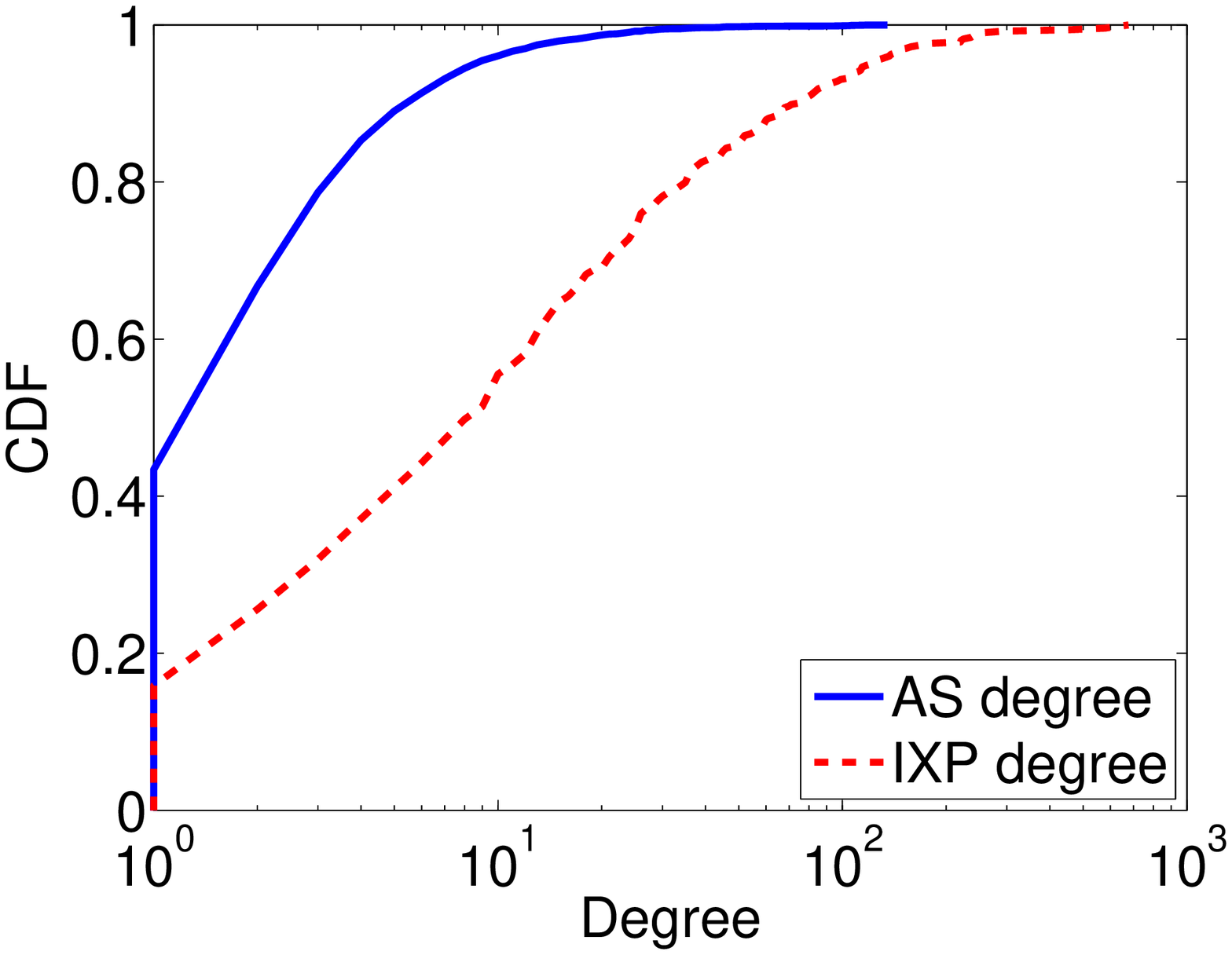}\label{fig:cdf-degrees}}
\subfigure[Types of IXP members]{\includegraphics[scale=0.23]{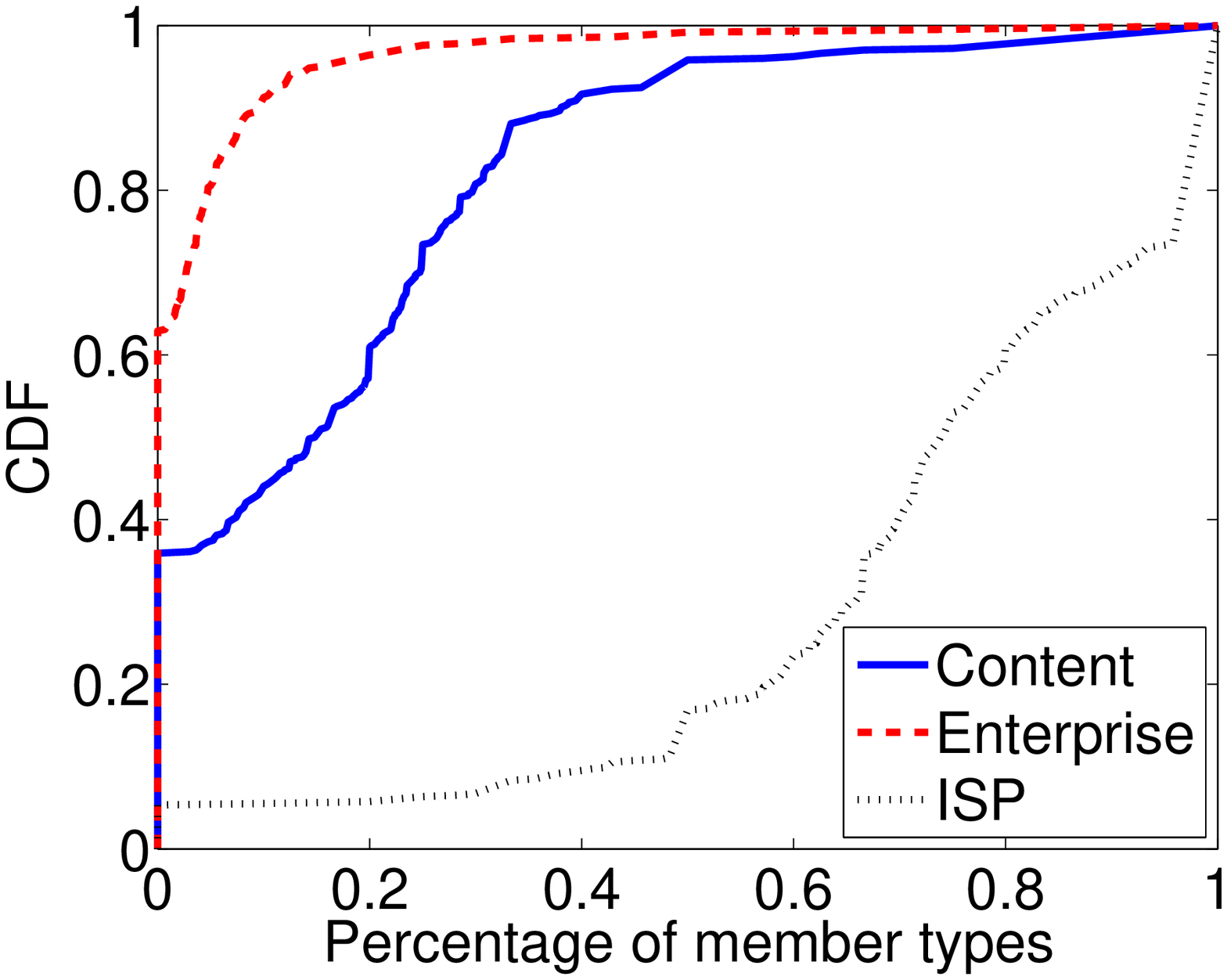}\label{fig:cdf-degrees-types}}
\caption{(a) CDF of the \textit{node degree} for IXP and AS nodes. (b) CDF of the \textit{percentage} of IXP members that belong to different \textit{AS-types}.}
\label{fig:members-memberships}
\end{figure}

\subsection{Graph Analysis and Applications}\label{sec:analysis}
From the collected datasets, we build the IXP bipartite graph, as well as, the projected graphs (IXP and AS multigraphs). We detect the main graph properties, and calculate complex network analysis metrics for the whole graph and each node set (IXPs and ASes). Due to space limitation, we present only a subset of the results, comprising the most important, closely-related to applications, and previously unexplored observations. For a clearer presentation, we group our findings in two categories: (i) members and memberships, and (ii) connectivity.

%%@Pavlos (+George): please add a readme here http://wired.ics.forth.gr/datasets/ . This url could also accommodate future datasets. 
\textbf{Members and Memberships.} The degree of a node (i.e., the number of edges connecting it to other nodes) in the IXP BG, depending on the node type, denotes either the \textit{number of AS-members} (for an IXP node), or the \textit{number of memberships to IXPs} (for an AS node). We present the degree distributions (namely, the CDFs, complementary distribution functions) for both node types in Fig~\ref{fig:cdf-degrees}. 

An AS connecting to an IXP has the possibility, with a \textit{single physical link}, to establish connections to other AS-members, and obtain access to their networks (peer ASes) and/or parts of the Internet (transit ASes). Hence the degree of an IXP indicates its size, importance, and attractiveness to ASes. As seen in Fig~\ref{fig:cdf-degrees}, the majority of IXPs have up to a few tens of members, while a small percentage ($7\%$ or $35$ IXPs) has more than $100$ members (the largest value of members in our dataset is $677$).

Using the AS-type information from~\cite{caida}, we calculate for each IXP the percentage of its members that belong to a specific AS-type (namely, content providers, enterprises, and ISPs). In Fig.~\ref{fig:cdf-degrees-types}, we present the corresponding CDFs. Some important observations are: (i) $36\%$ and $63\%$ of the IXPs do not have any member that is content provider and enterprise, respectively; (ii) in most IXPs, the majority of the members are ISPs, i.e., the percentage of members-ISPs is less than $50\%$ (x-axis $<0.5$) only for $16\%$ of the IXPs (y-axis $<0.16$).

%With respect to the degrees of the AS-nodes, the number of memberships to IXPs an AS has (i.e., its degree value) indicates the extent to which it can be \textit{directly accessible} by other ASes. An AS with presence to many IXPs can establish connections with many ASes, and thus reduce its transit costs (by using peering connections), or increase its revenue (offer transit to more customers), etc.

The degree of an AS-node, indicates the extent to which it can be \textit{directly accessible} by other ASes. An AS with presence to many IXPs (i.e., high degree) can establish connections with many ASes, and thus reduce its transit costs (by using peering connections), or increase its revenue (offer transit to more customers), etc. The AS-degree curve in Fig~\ref{fig:cdf-degrees}, shows that $43\%$ of the ASes are connected only to one IXP, which denotes that a large percentage of ASes mainly rely on transit agreements for Internet access. Only $4\%$ of the ASes (i.e., less than $200$ ASes) have memberships to more than $10$ IXPs, and $1\%$ (i.e., $6$ ASes) to more than $100$ IXPs. Towards obtaining a more detailed view of the AS-degree distribution, in Table~\ref{table:percentage-AS-types} we consider the node degree for each AS-type separately. As we can see, while only $19\%$ of \textit{all} ASes in our dataset are content providers (CDNs), in the subsets of ASes with degrees larger than $20$ and $30$, the corresponding percentages of CDNs increase significantly to $35\%$ and $64\%$. This clearly shows that CDNs tend to have presence in more IXPs, compared to other AS types, and reveals an interesting aspect of the dynamics (e.g., economic incentives) behind the inter-domain topology.

The above generic findings, could offer insights for the \textit{IXP selection strategy} of an AS desiring to set a new IXP membership (see also Section~\ref{sec:model-applications}). For instance, a small ISP, having a substantial amount of traffic to/from CDNs, should turn its attention to the list of IXPs with many CDN members (which are mostly large IXPs; cf. Table~\ref{table:percentage-AS-types}). On the other hand, a large ISP offering transit, is probably more interested in connections with other (small) ISPs; this indicates a market targeting small-size IXPs. Moreover, the IXP selection process could benefit from information about IP addresses belonging to each AS. However, an interesting finding in our datasets, is that there is \textit{no correlation} (correlation coefficient $<=0.1$) between the AS degree and network size (i.e., total announced IP prefixes).

\begin{figure}
\centering
\includegraphics[scale=0.23]{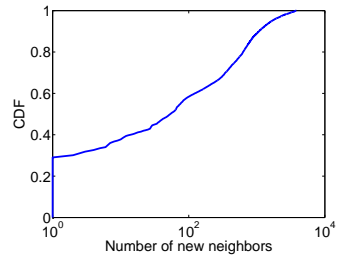}
\caption{CDF of the \textit{number of new neighbors} a node can access from a remote peering connection.}
\label{fig:similarity}
\end{figure}

Similarly, with respect to the \textit{remote peering} problem discussed in Section~\ref{sec:model-applications}, assume an $AS_{1}$ at $IXP_{1}$, and an $AS_{2}$ being member of both $IXP_{1}$ and $IXP_{2}$; $AS_{1}$ uses $AS_{2}$ as a ``tunnel'' to obtain access to the remote $IXP_{2}$ and peer with its members. A careful selection, among the ASes offering remote peering, can be done using the IXP BG, by calculating the number of new connections to ASes a ``tunnel'' can offer\footnote{Note that here, as AS can select among \textit{only the neighboring ASes}, while in the aforementioned case the selection is among \textit{all IXPs}.}. In Fig.~\ref{fig:similarity} we present this metric for all eligible AS-pairs in our dataset (i.e., AS-pairs colocated in at least one IXP).

\begin{table}[!h]
	\centering
	\caption{Percentage of AS types}
	\label{table:percentage-AS-types}
	%\resizebox{230pt}{23pt}{
		\begin{tabular}{|lccc|}
			\hline
			{\textbf{Subset of ASes}}&{\textbf{Content}}&{\textbf{Enterprise}}&{\textbf{ISP}}\\
			\hline
			{All ASes}				&{$19\%$}			&{$5\%$}					&{$69\%$}\\
			{ASes with degree $>20$}&{$35\%$}			&{$2\%$}					&{$63\%$}\\
			{ASes with degree $>30$}&{$64\%$}			&{$4\%$}					&{$32\%$}\\
			\hline
		\end{tabular}
		%}
	\end{table}

%\underline{Example Application:} An AS desiring to set a new IXP membership, can utilize the IXP BG to make a careful/optimal selection of the IXP. For instance, a small ISP, having a substantial amount of traffic to/from CDNs, should turn its attention to the list of IXPs with many CDN members (which are mostly large IXPs; cf. Table~\ref{table:percentage-AS-types}). On the other hand, a (large) ISP offering transit, is probably more interested in connections with other (small) ISPs; this indicates a market targeting small-size IXPs. Moreover, the IXP selection process could benefit from information about IP prefixes belonging to each AS. However, an interesting finding in our datasets, is that there is \textit{no correlation} (correlation coefficient $<=0.1$) between the network size of an AS (i.e., total announced IP prefixes) and other graph metrics.  Therefore, in order to use this information, a case-specific approach needs to be followed (e.g., considering traffic density to/from specific IP addresses).

\textbf{Connectivity.} We now proceed to study the connectivity between nodes in the Internet%(i.e., how close they are, how many direct connections or paths exist between them, etc.)
. We first calculate the shortest paths in the IXP BG between each AS-pair. Note that in the BG a path from a node $AS_{1}$ to a node $AS_{2}$ crosses at least one IXP. Table~\ref{table:AS-shortest-paths} gives the percentage of paths that cross a certain number of IXPs. Most paths ($\sim83\%$) cross exactly $2$ IXPs, while less than $2\%$ of the paths cross more than $3$ IXPs. Here, we need to stress that these paths do \textit{not} coincide always with the existing routing paths (e.g., determined by BGP) in the Internet, mainly due to restrictions imposed by peering policies, existence of private links, etc. However, they reveal the capabilities (under the given topology) for inter-domain routing solutions, including: failure or emergency cases management, novel routing services and economic models, etc.,~\cite{CXP,Hu-Measurement-ID-Routing-Diversity-2012}.

%\vspace{-0.2cm}
\begin{table}[!h]
\centering
\caption{Number of IXPs crossed by AS-to-AS shortest paths.}
\label{table:AS-shortest-paths}
%\resizebox{239pt}{13pt}{
\begin{tabular}{|l|ccccc|}
\hline
{Number of IXPs}		&{$1$}		&{$2$}		&{$3$}		&{$4$}		&{$5$}\\
\hline
{Percentage of paths}&{$7.9\%$}	&{$83.3\%$}	&{$8.6\%$}	&{$0.16\%$}	&{$0.001\%$}\\
\hline
\end{tabular}
%}
\end{table}

Moreover, in Table~\ref{table:edge-multiplicity} we calculate (from the IXP or AS multigraphs) the \textit{edge multiplicity} between node pairs of the same type, i.e., AS-pairs or IXP-pairs. In the case of IXP-pairs, edge multiplicity represents the number of common members (ASes). Since these members are mainly ISPs, they can offer connectivity to IXPs (similarly to their customers), e.g., to enable services hosted at IXP locations~\cite{CXP,Gupta-SDX-CCR-2014,decix-blackholing}. The edge multiplicity metric, quantifies the possible \textit{direct} connectivity between IXPs. As seen in Table~\ref{table:edge-multiplicity}, direct connectivity ($>=1$) can be achieved in a significant percentage of IXP-pairs ($20\%$, i.e., $25k$ IXP-pairs). In the case of AS-pairs, edge multiplicity represents the number of their common memberships in IXPs. Being connected in more than one points (IXPs), allows ASes to control better their inbound/outbound data traffic, by applying \textit{traffic engineering} techniques~\cite{Gupta-SDX-CCR-2014}. Table~\ref{table:edge-multiplicity} shows that $8\%$ of AS-pairs (i.e., $\sim800k$ pairs) have common presence in at least one IXP. Moreover, around $40\%$ of these pairs ($2.2\%$ of the total AS-pairs) have presence to more than one IXP.

\begin{table}[!h]
\centering
\caption{Edge multiplicity of IXP- and AS- pairs.}
\label{table:edge-multiplicity}
\resizebox{230pt}{17pt}{
\begin{tabular}{|l|cccccc|}
\hline
{Edge Multiplicity}		&{$0$}		&{$1$}		&{$2$}		&{$3$}		&{$4$}		&{$>=5$}\\
\hline
{\hspace{-0.1cm}Percentage of IXP-pairs\hspace{-0.15cm}}		&{$80.2\%$}	&{$7.9\%$}	&{$4.7\%$}	&{$2.3\%$}	&{$1.2\%$}	&{$3.7\%$}\\
{\hspace{-0.1cm}Percentage of AS-pais\hspace{-0.15cm}}		&{$92\%$}	&{$5.8\%$}	&{$1.4\%$}	&{$0.4\%$}	&{$0.2\%$}	&{$0.2\%$}\\
\hline
\end{tabular}
}
\end{table}

%\underline{Example Application:} Internet economics (peering agreements, SLAs, etc.) are highly dynamic, keeping up with technological advances. A related and promising application example is \textit{remote peering}~\cite{}: assume an $AS_{1}$ at $IXP_{1}$, and an $AS_{2}$ being member at both $IXP_{1}$ and $IXP_{2}$; $AS_{1}$ uses $AS_{2}$ as a ``tunnel'' to obtain access to the remote $IXP_{2}$ and peer with its members. A connectivity metric that is useful in this case is the AS-pair \textit{similarity}, which quantifies the number/percentage of common (second order) neighbors between two ASes. High similarity, means that the involved ASes share many neighbors. Thus, the gains of remote peering are larger when for nodes with \textit{low similarity}. The similarity metric can be easily calculated in the IXP BG; we present \red{........ in Fig.~\ref{}}.

%% file: Preliminaries_pavlos.tex
The Internet topology has been studied from a number of different perspectives. The main modeling approaches focus on the AS-, router-, and PoP- topology.

\textbf{AS topology.} Among the first efforts to study the inter-domain Internet topology and map it in an AS-level graph, are~\cite{govindan1997analysis,faloutsos1999power}. In AS-models, an AS is represented as a single node, while edges connecting AS-pairs correspond to AS-relationships and routing policies~\cite{muhlbauer2007search}, eBGP sessions~\cite{routeviews}, etc. The construction of the AS-graph mainly relies on BGP and router-level measurement data. However, measurements often suffer from missing or inaccurate data due to issues related to insufficient vantage points \cite{roughan201110}, Multiple Origins ASes (MOAS) \cite{zhao2001analysis}, load balancing \cite{augustin2006avoiding}, etc.

\textbf{Router topology.} Router-level models~\cite{pansiot1998routes}%spring2002measuring
, represent the~Internet with finer granularity than an AS-graph. A router-graph denotes routers as nodes, and physical links between routers as edges. Initial studies of the router topology revealed some interesting properties (e.g, power-low degree distribution)~\cite{faloutsos1999power}. However, inconsistencies between some findings and the real Internet, led researchers to re-evaluate the measurement methods and detect shortcomings, e.g, IP aliasing,~\cite{pansiot1998routes,keys2010internet}. %Although alternative approaches were proposed~\cite{fabrikant2002heuristically}, they did not yield the expected results~\cite{Willinger-Internet-Topology-ebook-2013}.

\textbf{PoP topology.} The Point of Presence (PoP) models~\cite{knight2011internet,giotsasmapping} map the geographical location of the Internet physical infrastructure. Since the PoP topology does not change frequently~\cite{shavitt2010structural}, we can get a more comprehensive view~in~network design principles (scalability, robustness, etc.)~\cite{cisco2005}. However, like router-models, PoP-models also suffer from inaccuracy issues due to traceroute measurements limitations~\cite{shavitt2010structural}.

Recently, several works highlighted the importance of IXPs in the Internet connectivity and structure~\cite{ager2012anatomy,augustin2009ixps,CXP}. Also, IXP-based data appear more often in studies relating to the Internet topology~\cite{dhamdhere2010internet,gregori2011impact,giotsasmapping}. In contrast to our work, these studies use IXP data to increase the completeness of the AS~\cite{dhamdhere2010internet,gregori2011impact} or PoP~\cite{giotsasmapping} level graphs, whereas our IXP-based model offers an entirely different perspective for the Internet topology. We deem that viewing the Internet from its central points, can facilitate IXP-related research and provide further insights on the interplay between topology and its use cases.

%% file: Conclusion.tex
Motivated by the increasing role and interest in IXPs, in this work, we propose an IXP-based model for the inter-domain Internet topology. Our model, the IXP bipartite graph, can be \textit{easily} built from \textit{publicly available} data, and avoid shortcomings of measurement methods employed by existing models. We believe that the IXP BG can substantially (a)~complement previous modeling approaches in~revealing interesting characteristics of the Internet topology, and (b)~support new research and applications in the areas of inter-domain routing, IXP-related innovation, and network economics (e.g, remote peering, outsourcing middleboxes, traffic engineering).